\documentclass[letterpaper,twoside,12pt]{article}
\usepackage{amssymb}
\usepackage{amsmath}
\usepackage{latexsym}
\usepackage{verbatim}
\usepackage{graphicx}
\usepackage{epsfig}
\usepackage{stmaryrd}
\usepackage{rotating,graphicx}
\usepackage[english]{babel}
\usepackage{setspace}
\usepackage[english]{babel}
\usepackage[utf8]{inputenc}
\usepackage[T1]{fontenc}
\usepackage{lmodern}
\usepackage{pstricks}
\usepackage{cjhebrew}

\makeatletter
\def\@xfootnote[#1]{%
 \protected@xdef\@thefnmark{#1}%
 \@footnotemark\@footnotetext}
\makeatother
\begin{document}

\newcommand{\bra}[1]{\langle #1|}
\newcommand{\ket}[1]{|#1\rangle}
\newcommand{\braket}[2]{\langle #1|#2\rangle}
\begin{large}
\begin{center}
\textbf{{\Large A (not so?) novel explanation for the very special initial state of the universe}}
\end{center}
\end{large}

\begin{center}
Elias Okon 

\textit{Instituto de Investigaciones Filos\'{o}ficas, Universidad Nacional Aut\'{o}noma de M\'{e}xico\\
Circuito Maestro Mario de la Cueva s/n, Distrito Federal, 04510, Mexico}\\
\textit{E-mail:} \texttt{eokon@filosoficas.unam.mx} \\ [.45cm]

Daniel Sudarsky 

\textit{Instituto de Ciencias Nucleares, Universidad Nacional Aut\'{o}noma de M\'{e}xico\\
Apartado Postal 70-543, Distrito Federal, 04510, M\'{e}xico}\\
\textit{E-mail:} \texttt{sudarsky@nucleares.unam.mx} \\[.45cm]

\end{center}
We put forward a proposal that combines objective collapse models, developed in connection with quantum-foundational questions, with the so-called Weyl curvature hypothesis, introduced by Roger Penrose as an attempt to account for the very special initial state of the universe. In particular, we explain how a curvature dependence of the collapse rate in such models, an idea already shown to help in the context of black holes and information loss, could also offer a \emph{dynamical} justification for Penrose's conjecture.\\

\begin{flushright}
\textbf{Genesis 1:1-2} .\cjRL{hAy:tAh tohU wAbohU } ...\cjRL{b*:re'+siyt}
\end{flushright}
\section{Introduction}
\label{Int}
\onehalfspacing
The second law of thermodynamics has often been a source of perplexity. The basic puzzle is how such a time-asymmetric law can emerge from fundamental laws of nature, such as those of general relativity and quantum theory, which are essentially\footnote{With the exception of the very small CP violation in the Electroweak interaction.} time-symmetric. 
 
The most convincing solution, endorsed by Boltzmann, Einstein, Feynman, and Schrödinger, to name a few, is that the universe started in a state of extremely low entropy. Penrose conjectured in \cite{Pen:79} that this, in turn, arises from a constraint on the initial state of the universe, prescribing that the Weyl curvature, as described, for instance, trough the scalar
 $W = \sqrt { | G^{abcdefgh} W_{abcd}W_{efgh} |}$ (where $W_{abcd}$ is the Weyl tensor and $G^{abcdefgh}$ is a supermetric constructed with appropriate combinations of the spacetime metric), should vanish as the initial singularity is approached. Such constraint prevents, for instance, a universe filled with primordial white holes, but does not prevent the emergence of ordinary black holes as the universe evolves. Clearly, the law that Penrose is proposing is very different from standard physical laws, which govern the dynamics, rather than the initial conditions.
 The aim of this essay is to sketch a scenario in which Penrose's conjecture has a dynamical origin (Penrose has a different proposal for this \cite{ConformalCosmology}).
 
It is convenient to clarify at the onset the context in which we frame our discussion. Most people are convinced that, underlying General Relativity, there must be a more fundamental theory of quantum gravity that will, among other things, ``cure the singularities.'' The idea is that the classical characterization of spacetime is just an approximation, valid in a limited set of circumstances. In particular, the singularities exhibited by the former are expected to be replaced by a non-singular characterization, not susceptible to a description in the classical language (as an example of this within the Loop Quantum Gravity approach see \cite{AshtekarBojowald}). Of course, these quantum gravity aspects are usually expected to be relevant only at the Planck scale, i.e., where the scalar curvatures $R$ or $W$ are of the order of the Plank scale ($M_{P}^2$ or $M_{P}^4$ respectively). 
 
Before presenting the details of our proposal in section \ref{Weyl}, we have some setting up to do. In section \ref{semi} we describe the semiclassical framework we employ, in section \ref{col} we talk about dynamical reduction models and in section \ref{tog} we explain how these two ingredients come together. Finally, in section \ref{con} we present our conclusions.
\section{Effective Spacetime and semiclassical Einstein equations}
\label{semi}
Our proposal is framed in a context in which the full quantum gravity regime has been ``exited,'' so we will assume that classical notions of spacetime already make sense. However, we will also assume that such a regime allows for a quantum description of matter fields. Therefore, we will be working at the, so-called, semiclassical gravity regime, which, as we explained above, will be viewed as an approximate and effective description, valid in limited circumstances, of a more fundamental theory of quantum gravity. We see all this as analogous to a hydrodynamic description of fluids, which works well in many circumstances, but does not represent the behavior of the truly fundamental degrees of freedom (corresponding to molecules, atoms and, ultimately, electrons, quacks and gauge bosons). Therefore, just as the Navier-Stocks equation describing a fluid must break down, say, when an ocean wave breaks at the beach, we expect Einstein's semiclassical equations to break down in some situations, such as those associated with a collapse of the quantum state (in which the energy-momentum tensor is not conserved).
 
In order to explore these ideas in detail, we need some scheme to implement them formally. We consider an approach initially proposed in \cite{Alberto} in the context of the emergence of the primordial inhomogeneities in inflationary cosmology, \cite{Inflation-Collapse}. The staring point is the notion of a Semiclassical Self-consistent Configuration (SSC), defined, for the case of a single matter field, as follows:

\textbf{Definition}: The set $\lbrace g_{ab}(x),\hat{\varphi}(x), \hat{\pi}(x), {\cal H}, \vert \xi \rangle \in {\cal H}\rbrace $ is a SSC if and only if $\hat{\varphi}(x)$, $\hat{\pi}(x)$ and $ {\cal H}$ correspond a to quantum field theory for the field $\varphi(x)$, constructed over a spacetime with metric $g_{ab}(x)$, and the state $\vert\xi\rangle$ in $ {\cal H}$ is such that:
\begin{equation}\label{scc}
G_{ab}[g(x)]=8\pi G\langle\xi\vert \hat{T}_{ab}[g(x),\hat{\varphi}(x)]\vert\xi
\rangle ,
\end{equation}
where $\langle\xi| \hat{T}_{\mu\nu}[g(x),\hat{\varphi}(x)]|\xi\rangle$ stands for the expectation value in the state $\vert\xi\rangle$ of the renormalized energy-momentum tensor of the quantum matter field $\hat{\varphi}(x)$, constructed with the spacetime metric $ g_{ab}$.

This characterization is a natural description of the kind of semiclassical situations we described above and is, in a sense, a general relativistic generalization of the so-called Schr\"odinger-Newton system \cite{Schrodinger-Newton}. However, as argued in \cite{Page-Gligker}, one must be careful not to take this description as fundamental because such a position would lead either to an internal inconsistency or to a conflict between the theory and experiments. In particular, in situations where quantum mechanics would lead to a delocalization of a macroscopic mass source, either: i) one assumes that there is no collapse of the wave function, in which case the gravitational filed observed does not correspond to the expectation value in equation (\ref{scc}); or ii) one assumes that there is a collapse of the wave function, in which case the equation simply would be inconsistent as the RHS would have a nonvanishing divergence (while Bianchi's identity ensures the divergence-free nature of the LHS). At any rate, as we will see below, the SSC formulation will allow for a self-consistent incorporation of collapses for the state of the quantum fields representing the matter sector. The basic idea is for different SSC characterizations to hold on the past and future of a collapse hypersurface and for them to be suitably joined at the collapse hypersurface. We will discuss this further in section 4. 

\section{Bringing in foundational aspects of quantum theory}
\label{col}
Most discussions of the sort we are having do not include foundational questions and difficulties of quantum theory. Therefore, it might seem strange for us to do so. Of course, often, issues of that kind can be safely ignored. However, as John S. Bell clearly argued in \cite{Bell.Cos}, such a pragmatic attitude is not always acceptable, and in some contexts, like those pertaining to the interface of quantum theory and cosmology, is not even viable (see also \cite{Hartle,Shortcomings, Oko.Sud:14}).

The text-book interpretation of quantum mechanics crucially depends on, either, the existence of observers external to the studied system or on some kind of artificial quantum/classical cut (see \cite{measurement}). Therefore, if the system in question is the whole universe, such cuts or external observers are simply unavailable and the standard formalism becomes inapplicable. What we require in order to apply quantum theory in these scenarios is an observer independent interpretation of the theory. Promising approaches in this regard are Bohmian mechanics \cite{Gold} and objective collapse models \cite{GRW}.\footnote{Many World scenarios and the Consistent Histories approach also aim at constructing observer independent formalisms. However, they seem to be plagued, at least for now, by insurmountable problems (see e.g., \cite[sec. 4]{MW} and \cite{CH}).} The latter, add stochastic and nonlinear terms to the dynamical equation of the standard theory in order to explain, in a unified way, both the behavior of micro-systems and the absence of superpositions at the macro-level. 

In order to frame our proposal in terms of a concrete example, we offer a brief presentation of two of the simplest objective collapse models, known as GRW and CSL. The first one, introduced in \cite{GRW:86}, postulates a probability for each elementary particle to suffer, at random times distributed with mean frequency $\lambda_0$, sudden localization processes around appropriate positions. Such localizations are then shown to quickly destroy superpositions of well-localized macroscopic states with centers separated by distances greater than the localization scale.

The mathematical framework of CSL or Continuous Spontaneous Localization, \cite{CSL}, is that of standard quantum mechanics with a \emph{modified} quantum dynamical evolution, specified by two equations: i) a modified Schr\"odinger equation, whose solution is
 \begin{equation}\label{CSL1}
 { |\psi,t\rangle_w = \hat {\cal T}e^{-\int_{0}^{t}dt'\big[i\hat H+\frac{1}{4\lambda_{0}}[w(t')-2\lambda_{0}\hat A]^{2}\big]}|\psi,0\rangle,}
\end{equation}
 where $\hat {\cal T}$ is the time-ordering operator and $w(t)$ is a random, white noise-type classical function of time, and ii) a probability rule for $w(t)$ given by
 \begin{equation}\label{CSL2}
	 { PDw(t)\equiv{}_w \langle\psi,t|\psi,t\rangle_w \prod_{t_{i}=0}^{t}\frac{dw(t_{i})}{\sqrt{ 2\pi\lambda_{0}/dt}}}.
\end{equation}
The CSL dynamics guaranties that, in the long run, the state of the system will be an eigenstate of {$\hat A$}. Therefore, {\it it unifies the standard Schr\"odinger evolution and the corresponding measurement of the observable {$\hat A$}}. For non-relativistic quantum mechanics one sets {$\hat A = \hat {\vec X} $} with $\hat{\vec X}$ a suitably smeared version of the position operator. 

Both within GRW and CSL, with a suitable value of $\lambda_{0}$ (see below), one can treat the whole system, including the measurement apparatus, quantum mechanically and make the correct predictions. Therefore, they successfully addresses the measurement problem. Regarding the value of $\lambda_{0}$, it should be small enough in order to avoid conflict with tests of quantum mechanics in the domain of subatomic physics, and big enough in order to ensure a rapid localization of ``macroscopic objects.'' Such constraints can be accomplished with the suggested value of {$\lambda_{0} \sim 10^{-16} sec ^{-1}$}. One of the drawbacks of the early versions of objective collapse models, such as \cite{CSL,GRW:86}, is the issue of compatibility with relativity. Fortunately, there have been important recent advances in resolving this problem and, in fact, at least three relativistic versions have been proposed, \cite{Tumulka,Bedingham,Pearle}.

In \cite{Oko.Sud:14} we argued that objective collapse models offer an array of benefits regarding the resolution of long-standing problems in cosmology and quantum gravity. In particular, we explored applications of objective collapse theories to the origin of seeds of cosmic structure, the problem of time in quantum gravity and the information loss paradox. Regarding black holes, in \cite{Oko.Sud:14,Oko.Sud:15} we suggested that the information paradox could be dissolved by a dynamical reduction theory in which the collapse parameter increases with $W$, so as to account for all the information destruction required. For example, one could have
\begin{equation}
\label{Wd}
\lambda \left( W \right) = \lambda_0 \left[ 1+\left( \frac{W}{\mu} \right)^\gamma \right]
\end{equation}
with $\gamma \geq 1$ a constant and with $\mu$ providing an appropriate scale. The idea is that such a $W$-dependence of the collapse parameter would guaranty all of the information encoded in the initial quantum state of the matter fields to be rapidly erased as the black hole singularity is approached. A successful application of these ideas, in the context of a two-dimensional model, was carried out in \cite{Sujoy} (non-relativistic objective collapse model) and \cite{Bed} (relativistic objective collapse model). In both works it is showed that the complete evaporation of the black hole, via Hawking radiation, does not lead to a paradox. 
\section{Collapse and the semiclassical spacetime context}
\label{tog}
Next we consider the implementation of an objective collapse model within the SSC scheme. For simplicity, we will consider a model with discrete collapses, akin to GRW. In order to add spontaneous jumps to the unitary evolution of the quantum field, given an SSC we denote as SSC1 (the situation before the collapse), the theory must randomly select i) a spacelike hypersurface $\Sigma_{C}$, on which the collapse takes place, and ii) the collapsed quantum state. Then, we must construct a new SSC, which we call SSC2 (the situation after the collapse), and join the two SSC's in order to generate a ``global spacetime.''

Again, we should think of the above scheme in analogy to an effective description of a fluid where, ``instantaneously,'' the Navier-Stokes equations do not hold. Consider once more an ocean wave breaking at the beach. Before the wave breaks, the situation is describable by Navier-Stokes equations, and the same is true well-after the brake when the surface becomes sufficiently smooth. However, the breakdown of the wave is not susceptible to a fluid description. Now let us take the limit in which the duration of the intermediate, non-fluid, regime tends to zero. In that case, we will have two regimes that are susceptible to a fluid description, but which are joined at an {\it instantaneous breaking of the wave}, to be identified in the analogy with the spacelike hypersurface $\Sigma_{C}$, which joins SSC1 and SSC2. 
 
The detailed procedure for the implementation of the gluing of SSC1 and SSC2 involves some subtleties. Suppose, for example, that we have an SSC1 and a (randomly selected) hypersurface $\Sigma_{C}$ in which the next GRW-type collapse is supposed to occur. Then, the state on $\Sigma_{C}$ will jump into a new state of the Hilbert space of SSC1. However, since the spacetime of SSC2 will have a different metric, the corresponding quantum field theory construction will, in general, differ from that of SSC1 and, thus, would involve a new Hilbert space $ {\cal H}_2 \not= {\cal H}_1$. Furthermore, the condition provided by equation (\ref{scc}) is not enough in order to uniquely determine the spacetime of SSC2.
 
All these issues have been resolved in \cite{Alberto} as follows: given SSC1, the dynamical collapse theory selects the hypesrurface of collapse $\Sigma_C$ and a \emph{tentative} or \emph{target} post-collapse state $ \vert \chi^{t} \rangle \in {\cal H}_1$. As we will see below, such state will be used in order to determine $\vert\xi ^{(2)}\rangle $, the actual state of SSC2. The SSC2 construction will be required to posses an hypersurface isometric to $\Sigma_C$, which will serve as the hypersurface where the two spacetimes will be joined. The requirement that the metrics induced on $\Sigma_C$ from the two spacetime metrics $g_{ab}^{(1)} $ and $g_{ab}^{(2)}$ coincide can be then seen as analogous to the Israel matching conditions for infinitely thin time-like shells. The actual state of the SSC2 construction will then be an element $ \vert\xi ^{(2)}\rangle $ of $ {\cal H}_2$ such that, on $\Sigma_C$, we have
\begin{equation}\label{scc-joining}
\langle\chi^{t}\vert \hat{T}^{(1)}_{ab}[g(x),\hat{\varphi}(x)]\vert\chi^{t}
\rangle =\langle\xi^{(2)}\vert \hat{T}^{(2)}_{ab}[g(x),\hat{\varphi}(x)]\vert\xi ^{(2)}\rangle ,
\end{equation}
where $\hat{T}^{(1)} $ and $\hat{T}^{(2)} $ are the renormalized energy-momentum tensors of the two SSC constructions, which depend on the corresponding spacetime metrics and the corresponding field theory constructions.
 
In \cite{Alberto} we have shown how these two additional requirements serve to specify the construction of SSC2, and thus, that of the complete spacetime which results from the gluing of the two constructions. It is also clear that Einstein's equations hold within the region corresponding to an individual SSC but will not hold along the gluing or collapse hypersurface $\Sigma_C$.

Generalizing this to the case of a multiplicity of collapses is straightforward if the dynamical collapse theory specifies the multiple collapse hypersurfaces. The only issue that might need further consideration is the fact that, unlike in a conventional quantum field theory, in which $\langle\xi_\Sigma \vert \hat{T}_{ab}(x) \vert\xi_\Sigma \rangle$ is independent of the choice of hypersurface $\Sigma$, in collapse theories such an expression, which is crucial in defining an SSC, does depend on $\Sigma$. That is because collapses between two hypersurfaces (that coincide in $x$) may affect its value. An attractive way out, proposed in \cite{RevMat}, is to specify that the hypersurface one must use in order to calculate $\langle\xi_\Sigma \vert \hat{T}_{ab}(x) \vert\xi_\Sigma \rangle$ is the past light cone of the point $x$. Finally, passing to a theory involving continuous collapses such as CSL would involve a limiting procedure which, at this point, we do not see as representing any particular difficulty in principle, although, clearly, any implementation of a concrete realization will represent a formidable task. 
\section{Collapse theories and the Weyl curvature conjecture}
\label{Weyl}
At last let us consider, within the general framework previously described, including a $W$-dependence of the collapse parameter such as equation (\ref{Wd}), the situation corresponding to the very early universe. We restrict ourselves to a post-planckian regime, in which the metric already provides a good characterization of spacetime. 

Assume that this early universe is characterized by wildly varying, generically high, values of curvature. Then, the collapse rate $\lambda (W)$ will be very large and the evolution will be dominated by the stochastic part of the modified quantum theory (i.e., the non-standard term in equation (\ref{CSL1}), with $\lambda_0$ substituted by $\lambda (W)$). As a result, the evolution of the matter fields and, through equation (\ref{scc}), of the geometry, will be extremely stochastic. This kind of evolution will continue until, by mere chance, the system obtains a small value of $W$. At such point, which can occur only when the spacetime is relatively smooth and the value of $R$ is rather uniform, the evolution will settle into the Hamiltonian dominated regime (i.e., the standard part of equation (\ref{CSL1})). That is, the evolution will settle into an almost purely-unitary regime only when the state of the matter fields is associated with an almost constant value of the density, a corresponding almost constant value of $R$ and a very small value of $W$. Of course, such scenario, with an extremely small value of $W$, precisely corresponds to Penrose's Weyl curvature hypothesis. Therefore, our scheme provides a dynamical justification for what Penrose introduced as a constraint on initial conditions in order to account for the very special initial state of the universe.

At this point we might stop to notice that the objective collapse dynamics of, e.g., equations (\ref{CSL1}) and (\ref{CSL2}) is in fact time reversal \emph{non}-invariant. In order to see this within CSL, we note that systems always evolve, into the future but not the past, towards eigenstates of the operator $\hat A$. However, the puzzle we started with, regarding the second law of thermodynamics, involved the fact that the laws of nature are time-symmetric. Why, then, should one go through all this trouble regarding the Weyl tensor? Well, as has been argued in \cite{A1}, collapse theories are, in fact, capable to explain why entropy is overwhelmingly likely to increase toward the future, but no toward the past. However, as has been argued in \cite{craig}, this does \emph{not} mean that such theories remove the need to additionally postulate that the universe started in a state of extremely low entropy. That is because, it is one thing to explain why, given a present nonequilibrium state, later states should have higher entropy (without implying that earlier states must have higher entropy too), and another to explain how the universe ever got to such a special nonequilibrium state in the first place. Our proposal, goes beyond that and offers a unified explanation of both these aspects; and its advantages do not end there.
 
After a small value of $W$ is randomly achieved, the stochastic part of the evolution becomes sub-dominant and the universe finds itself in the appropriate conditions for the onset of inflation. That process further flattens the spatial geometry and leads to the standard inflationary story, which successfully accounts for the main features of our present universe. The role of the collapse then, in general, becomes very small, but does not disappear completely. The minimal, non-zero, value of $\lambda(W \approx 0)\approx \lambda_0$ causes the small departures from homogeneity and isotropy, unexplainable within the standard approach, that eventually evolve into large-scale structure, \cite{Inflation-Collapse}. The collapse dynamics also continues to play the crucial role it was originally design for, i.e., the suppression of superpositions of well-localized ordinary macroscopic objects. The regime where $W$ is large is eventually reawakened in connection with the deep interior of black holes. This, as was argued in \cite{Oko.Sud:14,Oko.Sud:15, Sujoy}, accounts for the non-paradoxical erasing of information in the Hawking evaporation of such objects.
\section{Conclusions}
\label{con}
We have shown that objective collapse theories with a $W$-dependence of the collapse parameter offer a succinct account for a conundrum associated with the generalized second law of thermodynamics (i.e., the version which includes contribution from black holes to the overall entropy of a system). In particular, we have explained how such models provide a dynamical justification for the very special initial state of the universe. There is, of course, much to be done in order to complete this proposal. Still, we find it very encouraging that, by taking seriously modifications of quantum theory proposed in connection to foundational problems, this and other open issues in physics seem to find a solution. 

According to our proposal, the universe started in an arbitrary state, with a rather large value of the Weyl curvature. Then, due to the strong dependence of the constant determining the collapse strength on $W$, the dynamics where extremely chaotic. Such a chaotic behavior remained until, just by chance, the universe landed in a state with a small value of $W$. From then on, the standard evolution, dominated by the Hamiltonian term, prevailed, leading to a standard inflationary/big bang cosmology. It is hard not to notice the ``poetic similarity'' between the above picture and the description of the creation provided by the Book of Genesis: {\it In the beginning... there was chaos}. Perhaps, our idea is no so ``novel'' after all.

\section*{Acknowledgments}
We acknowledge partial financial support from DGAPA-UNAM project IG100316. DS was further supported by CONACyT project 101712 .

\end{document}